\documentclass[12pt,english]{article}
\usepackage[english]{babel}
\usepackage{natbib}
\usepackage[utf8]{inputenc}
\usepackage[lmargin=3.0cm,rmargin=2.5cm,tmargin=2.5cm,bmargin=2.5cm]{geometry}
\usepackage{graphicx}
\usepackage{url}
\usepackage{indentfirst}
\usepackage[pdfborder={1 1 1}]{hyperref} 
\hypersetup{
    bookmarks=true,         
    pdftoolbar=true,        
    pdfmenubar=true,        
    pdffitwindow=false,     
    pdfnewwindow=true,      
    colorlinks=false,       
}

\usepackage{subfig}
\usepackage{wrapfig}
\usepackage{amssymb}
\usepackage{t1enc}
\usepackage{textcomp}
\newcommand{\captionfonts}{\small}

\makeatletter 
\long\def\@makecaption#1#2{
  \vskip\abovecaptionskip
  \sbox\@tempboxa{{\captionfonts #1: #2}}
  \ifdim \wd\@tempboxa >\hsize
    {\captionfonts #1: #2\par}
  \else
    \hbox to\hsize{\hfil\box\@tempboxa\hfil}
  \fi
  \vskip\belowcaptionskip}
\makeatother

\title{Cosmic radiation drives quasi-periodic changes in the diversity of
siliceous marine microplankton} 

\author{Péter Ozsvárt,$^{1\ast}$ Emma Kun,$^{2,3}$\\
{$^{1}$HUN-REN-MTM-ELTE, Research Group for Paleontology,}\\
{PO BOX 137, 1431 Budapest, Hungary}\\
{$^{2}$Astronomical Institute, Faculty for Physics and Astronomy}\\
{Ruhr University Bochum, 44780 Bochum, Germany}\\
{$^{3}$Theoretical Physics IV: Plasma-Astroparticle Physics,}\\
{Faculty for Physics and Astronomy, Ruhr University Bochum,}\\
{Universitatstrasse 150, 44780 Bochum, Germany}\\
{$^\ast$To whom correspondence should be addressed;}\\
{E-mail: ozsvart.peter@nhmus.hu}
}

\begin{document} 

\maketitle 

\begin{abstract}
Radiolarians are significant contributors to the oceanic primary productivity and the global silica cycle in the last 500 Myr. Their diversity throughout the Phanerozoic shows periodic fluctuations. We identify a possible abiotic candidate for driving these patterns which seems to potentially influence radiolarian diversity changes during this period at a significance level of $\sim 2.2 \sigma$. Our finding suggests a significant correlation between the origination of new radiolaria species and maximum excursions of the Solar system from the Galactic plane, where the magnetic shielding of cosmic rays is expected to be weaker. We connect the particularly strong radiolaria blooming during the Middle Triassic to the so-called Mesozoic dipole-low of the geomagnetic field, which was in its deepest state when radiolarias were blooming. According to the scenario, high-energy cosmic rays presumably implied particular damage to the DNA during the maximum excursions which may trigger large chromosomal abnormalities leading to the appearance of a large number of new genera and species during these periods.
\end{abstract}

\section{Introduction}

Radiolarians are marine planktonic protists with complex, unique and diverse siliceous skeletons. They are one of Earth's oldest representatives of planktonic life and play a crucial role in the marine ecosystems from the Early Cambrian \citep{1} or Precambrian \citep{2} to the Recent as a dominant element in primary producers. Radiolarians are the most important organisms involved in the formation of biogenic siliceous sedimentary rocks (radiolarite), which are especially common worldwide from roughly mid-late Carboniferous to early Cretaceous deposits dominantly in pelagic deep oceanic basins. 

According to the massive paleontological record, the evolutionary history of radiolarians across the Phanerozoic presents repeated patterns of mass extinctions and rapid blooming and origination events \citep{3}. Relatively little information is available on the changes of radiolarian diversity during the Phanerozoic which may have been triggered by abiotic environmental factors such as repetitive volcanic activities, monsoon-driven upwelling activity \citep{4,5}, or simple thermic and chemical changes in the ocean surface waters. However, it is also possible that these diversity changes are regulated by biotic interactions such as natural selection, competition or other essential biotic effects. The radiolarian extinction events have been extensively studied over the past decades \citep{6,7}, however, significantly less research has focused on the precise causes of the originations or the diversification increases.

Based on the available data, the appearances of many new radiolaria species or genera occurred during relatively short time intervals, especially in the early Mesozoic period (Middle Triassic), although these sudden repeated blooming events have been also documented from the Middle Ordovician to the Early Miocene \citep{3}. Rapid blooms and accelerated evolutionary trends characterize the evolutionary history of radiolarians during those marked periods (see Fig \ref{fig1}.), which is generally followed by a long period of evolutionary equilibrium. Although the precise mechanisms and causes of the rapid origination events of radiolarians are not yet clear, our study may reveal a more complex pattern of biotic and abiotic dynamics than has previously been appreciated. 

Many studies show that extraterrestrial events could affect the biosphere in various ways, e.g. during meteorite events, due to changes in the cosmic radiation or even due to neutron star mergers \citep[e.g.][]{13,2016AAS...22721101M,2019Natur.569...85B,17}. The unique ratio of Potassium-40 ($^{40}$K) isotope to the $^{39}$K and $^{41}$K isotopes found in meteorites from the outer Solar system suggest the supernova origination of the their Potassium content \citep{doi:10.1126/science.abn1783}. All three Potassium isotopes are primarily produced in core-collapse explosions of massive stars (type II supernovae), but $^{40}$K is produced by more nuclear reaction pathways than the other two isotopes. The meteorites inherited nucleosynthetic anomalies of $^{40}$K produced in supernovae, and since this isotope with a long enough half-life of $1.25\times 10^9$ years is essentially frozen in these small rocky bodies, the cosmic radiation can bypass the atmosphere without interacting with it and contaminate the solid and water surfaces of Earth with a high dose of ionizing cosmic radiation (possibly in the form of $\beta$-decay). These studies are usually connected to extinction events \citep[e.g.][]{2020PNAS..11721008F}. In our scenario, the quasi-periodic appearance of radiolaria bloomings, enrichment of simple lifeforms in quality and quantity, is triggered by changes of the high-energy cosmic-ray flux in Galactic events.  The cosmic radiation is produced by electrically charged, ionizing particles, which originate from our Solar System (mainly from the Sun, at keV, MeV energies) or Galactic and extragalactic sources (MeV–EeV energies). The role of cosmic rays is important in galactic dynamics, they heat the interstellar gas, influence star formation and interact with magnetic fields. The observed energy spectrum of cosmic rays is dominantly power-law shaped, such that its energy axis spans about 12 orders of magnitude from 0.1 GeV to slightly more than $10^{11}$ GeV, and the flux axis spans 32 orders of magnitude (in particle/$m^2$/sr/s/GeV units), from 4 particles/$cm^2$/s to 1 particle/$km^2$/100 years \citep{18,19}. According to composition measurements, the spectrum is strongly proton-dominated up to $\sim100$ GeV \citep[e.g.][]{20}. The characteristic features of the spectrum are the ”first knee” at 3 PeV, the ”second knee” at 100 PeV, the ”ankle” at 8 EeV and the ”toe” at 60 EeV \citep[e.g.][]{21}. The transition between galactic and extragalactic origin occurs at about $10^9$ GeV, around the ankle.

The cosmic rays may trigger large chromosomal abnormalities and large-scale changes in chromosome structure that can affect the functioning of numerous genes, resulting in major phenotypic consequences \citep{30}. Genome changes caused by cosmic rays (by creating e.g. solvated electrons) could have led to abrupt changes in the external morphological characteristics of radiolarians that we see in each bloom period. 

The paper is organized as follows. In Section \ref{sec:radiolarias} we introduce the construction of our radiolaria genera database and reveal the particularly strong radiolaria blooming in the Triassic age. In Section \ref{sec:cosmicrays} we show the cosmic-ray variation during the oscillatory motion of the Solar system in the Galaxy. In Section \ref{sec:correlation} we correlate the quasi-periodic appearance of radiolaria bloomings with the vertical oscillatory motion of the Solar system about the Galactic plane. In Section \ref{sec:discussion}, we discuss how the cosmic radiation affects the DNA of simple lifeforms like the radiolarias. We also connect the particularly strong radiolaria blooming in the Triassic age to the so-called Mesozoic dipole low that was in its deepest state exactly when the Triassic radiolarias were blooming.

\section{Radiolaria bloomings}
\label{sec:radiolarias}
\subsection{Construction of the radiolaria genera database}
 
Estimating radiolarian diversity during the past 500 million years is a challenge, due to the heterogeneity of published data. The available databases (e.g. PBDB, Neptune, etc.) contain little or no information for the estimation of radiolaria diversity for the entire Phanerozoic. We focused on well-documented available radiolaria datasets of new Catalogs of Paleozoic, Mesozoic and Cenozoic radiolarians \citep{8,9,10} by reconstructing temporal patterns of genera and species originations. Unfortunately, this method has not been using sampling standardization (we do not have any information about the sample sizes), thanks to the highly confusing data reporting of previous publications, therefore the direct measurement of true diversity is rather uncertain. All of these catalogs have been created by immense efforts to represent the current state of the taxonomy and stratigraphic ranges of all known radiolarian genera, by the consensus of three active working teams. These radiolaria databases have allowed us to analyze more complete origination and biodiversity time series than before. 

\begin{figure}
    \centering
    \includegraphics[width=1.0\textwidth]{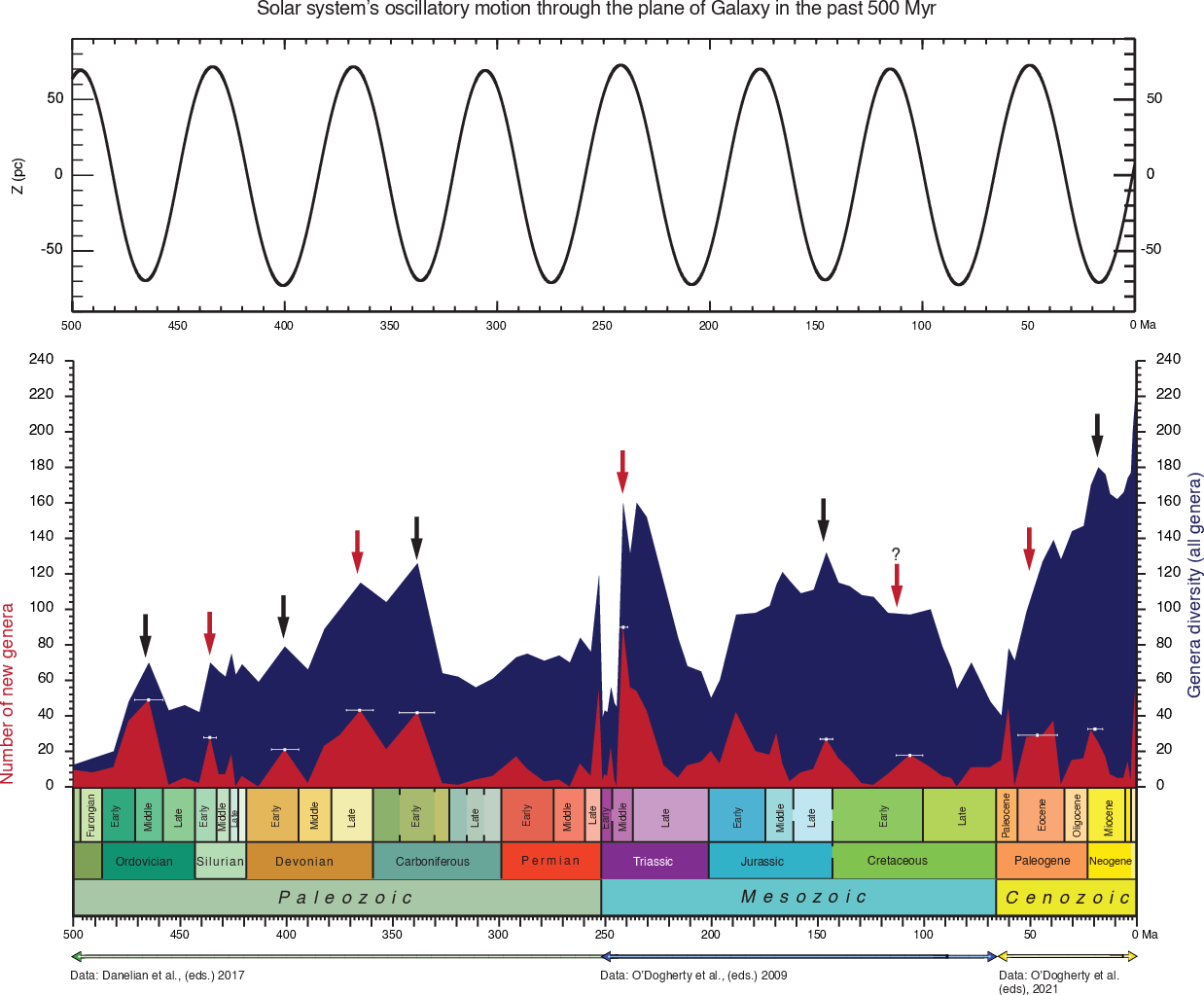}
    \caption{Top: Vertical motion of the Solar system about the Galactic plane, where $Z$ is the distance of the Solar system from the Galactic plane in parsec ($pc$). Bottom: Origination rates (red) and diversity (navy blue) for Phanerozoic radiolaria genera. The position of red arrows marks the maximum excursion of the Solar system from the Galactic plane towards North, while the black arrows show maximum excursion of the Solar system from the Galactic plane towards South.}
    \label{fig1}
\end{figure}

We obtained the first appearance datum (FAD) of radiolarians from these catalogs, reconstructed the origination curve and suggested a new pattern for the diversity curve across the entire Phanerozoic (Fig. \ref{fig1}). To minimize biases related to inconsistencies in sampling density in time, the raw data were smoothed and a curve was fitted onto the discrete data series. The origination records exhibit several steps and peaks that reflect on short episodes of appearances of new radiolaria genera and species. The first major origination event occurred in the latter mid-Ordovician (Darriwilian), exhibiting the highest peak in the entire Paleozoic, followed by several smaller peaks from the Silurian to the Late Devonian. We find a significant increase in origination in the Early Carboniferous (Visean) followed by two minor blooms during the Permian, and peaked with the end-Permian (Changhsingian), right before the greatest extinction event (P/T crisis) in the history of the biosphere. The largest number of new genera and species originations can be detected during the Middle to early Late Triassic interval in the entire Phanerozoic, while a smaller peak appears in the latest Triassic (latest Norian-Rhaetian) to Early Jurassic (Hettangian) period. During the entire Jurassic period, three major peaks can be detected, while in the Cretaceous there is a single significant bloom at the end of the Early Cretaceous (Albian). There are also three major increases in originations in the Cenozoic: the mid-Paleocene, the mid-Eocene, and the early Miocene. 

\subsection{The Triassic blooming}

Of all the origination events, the accelerated evolution of siliceous microplankton during the mid-Triassic period stands out, and we would like to illustrate the basic features of a blooming process by presenting it in more detail. This relatively short interval is marked by a huge diversification of all radiolaria groups, unprecedented morphological innovation and the appearance of a large number of very short-lived new species or mutants. Numerous new morphological characteristics of radiolarian tests and spines were recognized in all different groups from this interval \citep{11}. Typically, many new radiolarian species appearing in the individual Triassic blooms are only detectable in distinguished horizons but later completely disappear from the paleontological record \citep{12}. Middle Triassic radiolarians seem to exhibit a uniquely rapid evolutionary development at these distinct moments. The patterns of abiotic change in Middle Triassic radiolarians were explained by several quite generalized models, however, neither is fully supported by the available data. The traditional hypothesis is regarding the role of oceanic nutrient supply. There are currently two different models, the nutrient supply by upwelling \citep{4} and the monsoon-driven nutrient transport into the oceanic basins by river plume \citep{5}. In the first model, likely the breakup of the Pangea supercontinent and the partitioning of the Panthalassa oceanic basin contributed to the development of new oceanic circulation which might induce the up-welling of nutrients \citep{4}. Although this might have played an important role in the innovation and diversification of planktic organisms, it is more realistic that its primary role may have been in the spread of radiolarians. The monsoon-driven model, increasing the terrestrial weathering and transport of nutrients into the oceanic basin has been developed and applied primarily to the frequently formed Middle to Late Jurassic radiolarite sequences in western Tethys and proto–Atlantic Ocean \citep{4,5,13,14,15}, but it may have led to an increase in diversification but rather to an increase in productivity. 

These models mainly explain long-term changes in nutrient flux and their impact on productivity, while some alternative models might explain short-term diversification growth. In the simplest approach, increasing the supply of biologically key materials to surface water results in biological diversification. Volcanic activity, especially ash fall, might have played an important role in oceanic fertilization during the Middle Triassic. This would support the hypothesis that radiolarian diversification is a pulsating process, i.e. a short period of blooms followed by a long period of evolutionary equilibrium. Although the acidic volcanism took place continuously from the Scythian (Early Triassic) to the Norian (Late Triassic) in the entire Western Neotethys \citep{16}, especially the volcanic ash layers are quite frequently interbedded in the pelagic carbonates, although rich radiolarian assemblages only appear in a single selected horizon. These unique evolutionary innovations and accelerated rates of evolutionary changes might have been caused by direct environmental influences on the radiolarian genome like a strong cosmic flux, although the roles of abiotic and biotic drivers of this particularly important diversification event remain unclear so far.

\section{Cosmic-ray variation during the oscillatory motion of the Solar system in the Galaxy}
\label{sec:cosmicrays}

\begin{figure}
    \centering
    \includegraphics[width=1.0\textwidth]{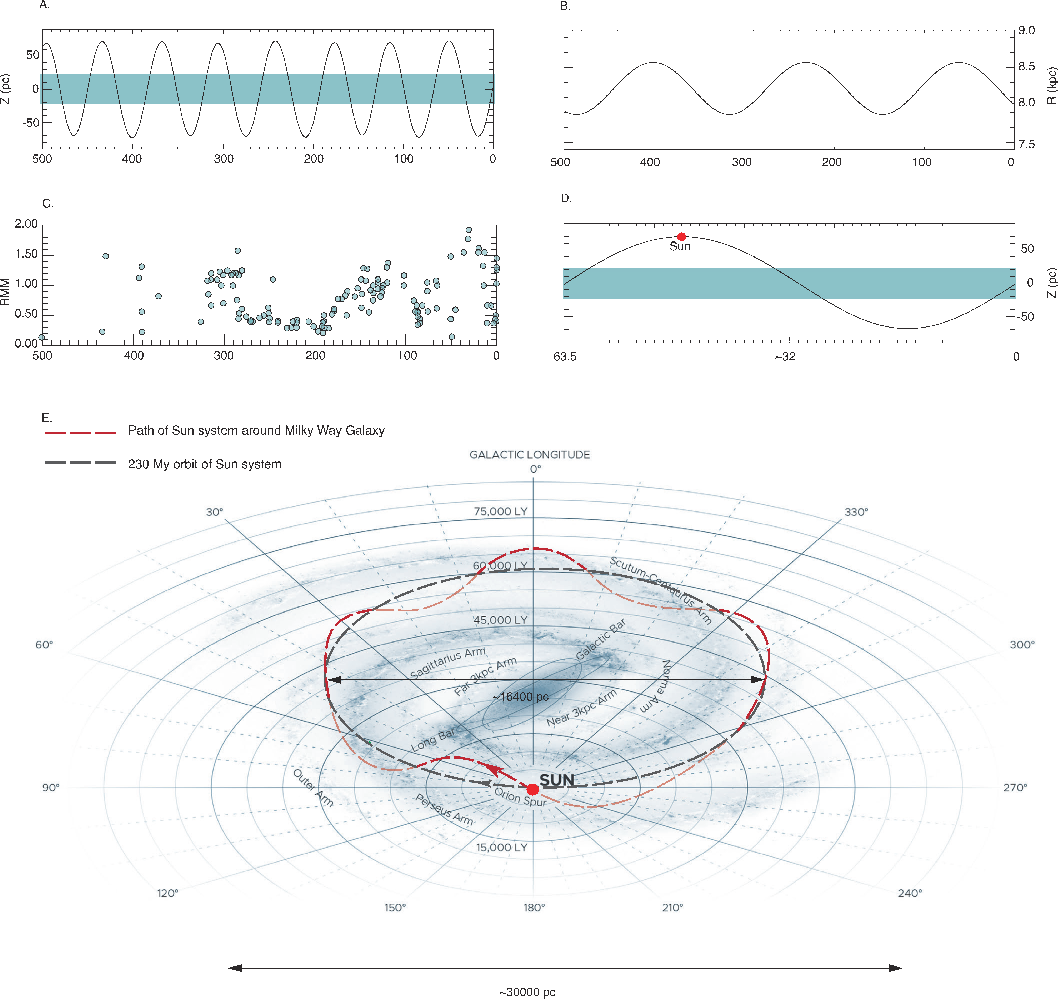}
    \caption{The Sun system’s oscillatory motion in our Galaxy. A: Solar system’s oscillatory motion through the plane of the galaxy in the past 500 Myr \citep{22}; B: Solar system’s distance from the Galactic center \citep{22}; C: Intensity of the magnetic field in the past 500 Myr \citep{34}; D: The period of the Solar system oscillation perpendicular to the Galactic disk $P_z = 63.5$ Myr and the maximum excursions of the Solar system from the Galactic plane, where the magnetic shielding of cosmic rays is expected to be strongest; E: Solar system moving in the Galaxy as viewed from above the Galactic plane (Credit: \url{solarsystem.nasa.gov/beyond}).}
    \label{fig2}
\end{figure}

Our Solar system is orbiting around the Galactic center with a period of about 230 million years (Fig. \ref{fig2}). The Solar system is not just simply in orbit, it oscillates vertically and radially in the Galactic disk (Fig. \ref{fig2} A, B, respectively), with a period of about 63 million years and with a vertical amplitude in the order of $\sim70$ pc \citep{22}. The Milky Way, an SBbc typed barred spiral galaxy, can be separated into disk, bulge, baryonic and non-baryonic halo components. The disk component consists of a young and an old thin disk (composed dominantly of gas, dust or young stars) and a thick disk (dominantly stars). Recent estimation of the vertical spatial structure of the Milky Way at the Solar radius using Gaia photometry and astrometry constrained the thin disc scale height 260 $\pm$ 26 pc (sys) and the thick disc as 693 $\pm$ 121 pc \citep{23}. This means our Solar system oscillates in the thin Galactic disk dense with stars, gas and dust, where a large flux of Galactic cosmic rays is expected e.g. due to frequent supernovae explosions, or the ionizing effect of hot, massive O-B stars. 

When the Solar System is located well within the Galactic plane, although the Galactic magnetic field is weak \citep[a few $\mu G$, e.g.][]{31,32}, its magnetic field can act as a shield protecting the biosphere from cosmic radiation thanks to the predominantly toroidal magnetic field of our Galaxy \citep{33}. We hypothesize that when the Solar system is maximally off north or south of the Galactic disk (Fig \ref{fig2}A), the biosphere is more exposed to cosmic radiation, and due to non-perfect DNA repairs, this exposure to ionizing radiation leads to mutations at the macro level (see a more detailed explanation in Section \ref{sec:discussion}). As a result, we eventually see remnants of radiolaria blooming events in these special time intervals.  Such periods of enhanced cosmic radiation could be identified directly based on geochemical evidences measuring the dose of isotopes with $\sim$ billion years of half life (e.g. $^{232}$Th -- $1.41\times 10^{10}$ yrs, $^{40}$K -- $1.25\times10^9$ yrs,  $^{238}$U -- $4.5\times10^9$ yrs, $^{235}$U -- $7\times10^8$ yrs). Finding the right isotopes with cosmic origin that can be separated from the terrestrial component is key, but potentially challenging due to e.g. volcanism, or monsoons. The unique ratio of K-isotopes produced in core-collapse supernovae and inherited by meteorites coming from the outer Solar system \citep{doi:10.1126/science.abn1783} is particularly interesting in this manner. 

\section{Correlation between the quasi-periodic radiolaria bloomings and Galactic events}
\label{sec:correlation}

In Fig. \ref{fig1}, the timing of the appearance of new radiolaria genera appears to follow the timing of the maximum displacements of the Solar system from the Galactic plane. We carried out a hypothesis test to decide if this connection is random, or could have a common origin. The null-hypothesis states the relationship between the appearance of new radiolaria genera and the Galactic events is merely coincidental, while the alternative hypothesis states the connection is not just the result of random processes, but it is real. First, we obtained the period of the Solar system oscillation perpendicular to the Galactic disk as $P_z = 63.5$ Myr \citep{22}. Then we randomized the initial phase of the Solar system periodicity $\phi_{0,rnd}$ between $\phi_0$ and $\phi_0 + 0.5$, where $\phi_0$ is the real initial phase of the oscillation, and calculated a new curve of the vertical oscillation assuming $\phi_{0,rnd}$ and $P_z$. After that, we selected time intervals from our radiolaria genera database during which the Solar system was maximally north or south of the Galactic disk ($T_1$, $T_2$, ..., $T_n$), according to the randomized oscillation. Then we summed up the number of new radiolaria genera in the n time intervals. This number, $X$, is the stochastic variable in our test. Repeating the process $10^5$ times, we get the discrete probability density function of $X$. Integrating this distribution from the observed value, $X_{obs} = 413$, to positive infinity, we get the probability of the connection between the number of new radiolaria genera and the maximum deflection of the Solar system from the Galactic plane being random as small as $P (X > X_{obs}) = 0.055$. This means we can reject the null-hypothesis at about $2\sigma$ significance level (assuming normal distribution). Since the time intervals have different widths, we repeated the same analysis by weighting the number of new genera with the time-width of the interval in which they appeared $(X_{obs}\sim60)$ genera/million years. Then the p-value decreased to $P (X > X_{obs}) = 0.027$ $(\sim 2.2 \sigma)$, hinting at the idea of a possible correlation between the appearance of new radiolaria genera and Galactic events. 

\section{Discussion and Summary}
\label{sec:discussion}

Our results suggest that one of the most direct external drivers of diversification \citep{17} might be the periodic chance in cosmic ray intensity during the vertical oscillatory motion of the Solar system about the Galactic plane. Although this change in cosmic radiation intensity is perhaps the most difficult to document in the sedimentary record, there are abundant shreds of evidence that it may have played an important role in explosive diversification. 

The high-energy cosmic rays interact with the upper part of the atmosphere of Earth, leading to the shower of secondary particles. A bulk of these particles are deflected into outer space by the magnetosphere and the heliosphere, but part of their flux is able to reach the surface of Earth. The muon component dominates in the secondary particles, contributing 85\% of the radiation dose from cosmic rays \citep{24}. The ozone depletion might have also played an important role in this process which increases the ultraviolet flux \citep{25}. Simulations indicate the proton showers with energies of $10^4-10^5$ GeV, at which energies protons dominate the observed cosmic-ray spectrum with the most possible origin of being Galactic \citep{21}, lose their energy within the first $\sim 10$ meters from the surface in the seawater \citep[66.2\% H, 33.1\% O, 0.7\% NaCl,][]{TSloan_2007}, where one expects a large amount of radiolarians. These high-energy protons interact with the water, leading to the excess of solvated electrons and free electrons in aqueous solutions, which come from the decay of muons created in the interaction of high-energy protons and water. Solvated electrons can potentially damage the DNA of living organisms \citep{Kumar2019}, introducing changes in the electronic structure of the biomolecules that presumably implied particular damage to the DNA and proteins of radiolarians. Enhanced levels of muon radiation from proton air showers also introduce potentially harmful effects on living organisms. The most common damage can be the double-strand break DNA, although there are various models that cells are able to repair damages quickly. If the high-energy cosmic-ray flux is increased, this repair mechanism can be inadequate which causes frequent gene mutations even for deep-sea organisms \citep{29}. In these morphological renewals, the actin-cytoskeleton might regulate the repair process, which is a dynamic network made up of actin polymers and associated actin binding proteins. Exposition of the biosphere to ionizing radiation might lead to the mutation of simple living systems such as radiolarians.

In Section \ref{sec:radiolarias}, we discussed the exceptional radiolaria blooming in the Middle Triassic and now we suggest an adiabatic reason of this. The Earth’s magnetic field is stronger than that of the Galaxy. The geomagnetic field has a fundamentally poloidal structure, the position of the poles and the strength of the magnetic field vary over the ages of the Earth’s history. During the Triassic period, the Earth’s magnetic field strength was at a deep minimum \citep{34} and experienced the so-called Mesozoic low (Fig. \ref{fig2}C). During the largest radiolarian bloom in the Middle to Late Triassic (Fig. \ref{fig1}), our Solar System was above the plane of our Galaxy. The extra exposure of the biosphere to cosmic radiation might made the Triassic age unique and the bloom of Middle Triassic radiolarians strong. 

The resulting p-value 0.027 (see Section \ref{sec:correlation}) hints the idea that variations in the flux of cosmic radiation due to the oscillatory motion of the Solar system might play an important role to induce the quasi-periodic diversity dynamics in siliceous marine microplankton across the entire Phanerozoic. However, it is also clear that not all maximum excursions resulted in strong blooms among radiolarians, although the exact reasons remain unclear. It seems plausible that far from the Galactic plane, where the magnetic shielding of the Galaxy is reduced, the enhanced flux of cosmic rays was able to reach Earth’s surface and could have affected the DNA of radiolarians, leading to the appearance of new genera and species. The resulting increase in diversity is controlled by natural selection, which has led to a slow evolutionary equilibration of radiolarians following the blooms. These patterns suggest that increasing cosmic rays may significantly influence marine ecosystems, and ultimately played an important role in the formation of several major radiolarian groups throughout the Phanerozoic. The quasi-periodic repetition in increasing diversity might also be seen among other groups of fossil organisms, although this has not yet been studied in detail. It will be important in the future to consider other terrestrial and cosmic mechanisms that might affect this picture and test other simple forms of life that could potentially be affected by variable flux of cosmic radiation.

\section*{Acknowledgments}
The authors thank Andr\'as T\'oth for fruitful discussions on the biological aspect of this work. E.K. thanks the Alexander von Humboldt Foundation for its Fellowship. 

\section*{Author contributions}
Conceptualization: PO, EK. Methodology: EK, OP. Visualization: PO, EK. Project administration: PO, EK. Writing – original draft: PO, EK. Writing – review and editing: PO, EK. Competing interests: Authors declare that they have no competing interests.

\end{document}